\documentclass[final,leqno,onefignum,onetabnum]{siamltex1213}

\usepackage{amsmath,amsfonts}
\usepackage{graphicx}
\usepackage{tikz}
\usepackage{framed}
\usepackage{mathtools}
\usepackage{siunitx}
\usepackage{color}

\DeclareSIUnit\molar{\mole\per\cubic\deci\metre}
\DeclareSIUnit\Molar{\textsc{m}}
\DeclareSIUnit\brate{\micro\Molar\tothe{-1}\second\tothe{-1}}
\DeclareSIUnit\ubrate{\second\tothe{-1}}

\newcommand{\ca}{Ca$^{2+}$}
\newcommand{\medcap}{\mathbin{\scalebox{1.5}{\ensuremath{\cap}}}}%

\begin{document}

\title{Particle-based Multiscale Modeling of\\ Calcium Puff
Dynamics\thanks{Submitted to {\it Multiscale Modeling and Simulation}
(\today). The research leading to these results has received 
funding from the European Research Council under the \textit{European
Community}'s Seventh Framework Programme 
({\it FP7/2007-2013})/ ERC {\it grant agreement} n$^o$  239870.}}
\author{Ulrich Dobramysl$^1$ 
\and Sten R\"{u}diger$^2$  
\and Radek Erban$^1$}
\def\today{April 1, 2015}

\maketitle

\footnotetext[1]{Mathematical Institute, University of Oxford, Radcliffe Observatory Quarter, Woodstock Road, Oxford, OX2 6GG, United Kingdom;
e-mails: dobramysl@maths.ox.ac.uk; erban@maths.ox.ac.uk.
Radek Erban would like to thank the Royal Society 
for a University Research Fellowship and the Leverhulme Trust 
for a Philip Leverhulme Prize.}
\footnotetext[2]{Institut f\"ur Physik, Humboldt-Universit\"at zu Berlin, 
12489 Berlin, Germany; e-mail: sten.ruediger@physik.hu-berlin.de.
Sten R\"{u}diger acknowledges support from the Deutsche 
Forschungsgemeinschaft (RU1660 and IRTG 1740).}

\slugger{mms}{xxxx}{xx}{x}{x--x}%slugger should be set to mms, siap, sicomp, sicon, sidma, sima, simax, sinum, siopt, sisc, or sirev

\begin{abstract}
  Intracellular calcium is regulated in part by the release of \ca{} ions from
  the endoplasmic reticulum via inositol-4,5-triphosphate receptor (IP$_3$R)
  channels (among other possibilities such as RyR and L-type calcium
  channels). The resulting dynamics are highly diverse, lead to local calcium
  ``puffs'' as well as global waves propagating through cells, as observed in
  {\it Xenopus} oocytes, neurons, and other cell types. Local fluctuations in the number
  of calcium ions play a crucial role in the onset of these features. Previous
  modeling studies of calcium puff dynamics stemming from IP$_3$R channels have
  predominantly focused on stochastic channel models coupled to deterministic
  diffusion of ions, thereby neglecting local fluctuations of the ion
  number. Tracking of individual ions is computationally difficult due to the
  scale separation in the \ca{} concentration when channels are in the open or
  closed states. In this paper, a spatial multiscale model for investigating of
  the dynamics of puffs is presented.  It couples Brownian motion (diffusion) of
  ions with a stochastic channel gating model. The model is used to analyze
  calcium puff statistics. Concentration time traces as well as channel state
  information are studied. We identify the regime in which puffs can be found
  and develop a mean-field theory to extract the boundary of this regime. Puffs
  are only possible when the time scale of channel inhibition is sufficiently
  large.  Implications for the understanding of puff generation and termination
  are discussed.
\end{abstract}

\markboth{ULRICH DOBRAMYSL, STEN R\"{U}DIGER AND RADEK ERBAN}{}



\section{Introduction}

Intracellular calcium plays a major role in many signaling pathways, and
regulates enzymatic activity, gene expression~\cite{Berridge2000}, and neural
activity \cite{zucker2002short,neher2008multiple}.  In order to control a
variety of cell functions, cells control the local cytoplasmic calcium
concentration via the exchange of \ca{} ions with the extracellular space and
the release of \ca{} ions through channels situated on the membrane of
reservoirs, such as the endoplasmic and the sarcoplasmic reticulum. Modulation
of the \ca{} release regulates muscle contraction, pathway cross-talk and
mitochondrial activity, and disruption of these processes is associated with
various diseases such as early-onset Alzheimer's~\cite{Berridge2011,Woods2012},
heart failure~\cite{Anderson2011,Anderson2014} as well as psychological
conditions such as bipolar disorder and
schizophrenia~\cite{Berridge2014}. Hence, detailed knowledge about the
underlying processes governing this signaling mechanism is required to allow
progress in our understanding of these diseases.

In this paper, we focus on the release of calcium ions from the endoplasmic
reticulum (ER) via inositol-4,5-triphosphate receptor (IP$_3$R) channels.  Upon
binding of \ca{} to binding sites on its cytosolic part, a channel opens and
calcium ions flow from the ER into the cytoplasm. Channels occur in clusters of
10 to 20 channels~\cite{Dickinson2012,Smith2009}. The opening of a single
channel usually triggers the release of \ca{} from other channels in the same
cluster due to the increased \ca{} concentration in their
vicinity~\cite{sherman2001asymptotic}.  This mechanism results in a highly
localized increase of the cytosolic calcium concentration. These ``puffs'' of
calcium ions have been detected and analyzed in experiments by using fluorescent
calcium buffers~\cite{Yao1995,Fraiman2006}.

Traditional modeling approaches of intracellular calcium dynamics are based on
deterministic macroscopic rate equations~\cite{Goldbeter1990,DeYoung1992},
however the intrinsically random, erratic nature of calcium signals in many
cells or cell domains necessitates an approach going beyond the deterministic
regime
\cite{zeng2010effect,franks2002complexity,holcman2005calcium,ruediger2014stochastic}. Progress
has been made by recognizing the importance of number fluctuations in the
binding to the channels~\cite{Swillens1999,Falcke2003} and using hybrid models,
where the deterministic calcium concentration is coupled to stochastic channel
binding models~\cite{rudiger2007hybrid}. Recently, it was found that local
fluctuations stemming from diffusive noise (i.e. noise originating from the
random movement of ions) have a crucial influence in calcium dynamics for
clusters of intracellular channels, particularly in the inter-puff waiting
time~\cite{Flegg2013} but also on single-channel equilibrium
behavior~\cite{wieder2015exact}. Stochastic effects were also investigated for
L-type calcium channels and RyR channels in the dyadic
cleft~\cite{Koh2006,Tanskanen2007,Hake2008}. Other studies consider a Langevin
equation governing the fraction of open channels in a
cluster~\cite{Wang2015}. However, tracking the exact diffusion of individual
ions in the complete computational domain is computationally intensive.

Here, for the first time, we apply spatial stochastic multiscale methods to
model the dynamics of calcium puffs including the release of \ca{} from IP$_3$R
channels and track individual ion positions in order to accurately incorporate
diffusive noise.  We take into account both activating and inhibitory channel
properties. We study the dynamics of the calcium concentration as a function of
the ion binding affinities and explore the regime where there exist puffs, as
well as the regime in which channels do not close after their first opening. The
paper is organized as follows: In the next Section~\ref{sec:model} we discuss
the spatial stochastic model for diffusion and channel gating used throughout
this study. Section~\ref{sec:multiscale} discusses the multiscale approach we
employ in order to reduce the computational effort required to track single ions
and hence make this study feasible. In Section~\ref{sec:results} we present
results on the statistics of puffs extracted from simulated time series data,
and study the transition between perpetually-open channel clusters and the
parameter regime in which puffs can be observed. In addition, we develop a
mean-field model for channel dynamics and use it to extract the boundary between
the regimes. Finally, we summarize our findings in
Section~\ref{sec:conclusions}.

\section{Spatial stochastic model for intracellular \ca{} release}
\label{sec:model}

The spatial extent of our computational model consists of the 
three-dimensional domain $\Omega$ which models a part of the 
intracellular space. \ca{} ions are able to undergo free 
diffusion in $\Omega$. They bind to and dissociate from binding 
sites on the channels, which are positioned on a small area of 
the domain boundary, corresponding to the membrane of the ER.
The domain geometry and boundary conditions are specified 
in Section~\ref{secgeometry}. In the following sub-sections 
we discuss the components of this model in detail. The parameter 
values used throughout this study can be found in Table~\ref{tab:parameter}.

\begin{table}
\centering
\begin{tabular}{|l|c|c|}
\hline
Parameter description\rule[-5pt]{0pt}{15pt} & Name & Value \\
\hline
\hline
\multicolumn{3}{|l|}{\rule[-5pt]{0pt}{15pt}%
  Parameter values from the literature} \\
\hline
\rule{0pt}{4mm}
~Free Ca$^{2+}$ diffusion constant~\cite{Allbritton1992} & $D$ & $\SI{220}{\micro\meter\tothe{2}\per\second}$ \\
\hline
\rule{0pt}{4mm}
~Cytoplasmic Ca$^{2+}$ concentration~\cite{Rudiger2010} & $c_0$ & $\SI{0.02}{\micro\Molar}$ \\
\hline
\rule{0pt}{4mm}
~Edge length of computational domain~\cite{Rudiger2010} & $L$ & $\SI{5}{\micro\meter}$ \\
\hline
\rule{0pt}{4mm}
~Rate of binding to activating site~\cite{Rudiger2010} & $a_a$ & $\SI{100}{\brate}$ \\
\hline
\rule{0pt}{4mm}
~Rate of unbinding from activating site~\cite{Rudiger2010} & $b_a$ & $\SI{20}{\ubrate}$ \\
\hline
\rule{0pt}{4mm}
~Open channel current~\cite{Bruno2010,Smith2009,Vais2010} & $I_C$ & $\SI{0.1}{\pico\ampere}$ \\
\hline
\rule{0pt}{4mm}
~Number of channels in cluster~\cite{Rudiger2010,Dickinson2012,Swillens1999} & $C$ & $\num{9}$ \\
\hline
\hline
\multicolumn{3}{|l|}{\rule[-5pt]{0pt}{15pt}%
  Chosen simulation parameters} \\
\hline
\rule{0pt}{4mm}
~Spacing between channels in cluster & $\ell$ & $\SI{0.15}{\micro\meter}$ \\
\hline
\rule{0pt}{4mm}
~Rate of binding to inhibitory site & $a_i$ & varies \\
\hline
\rule{0pt}{4mm}
~Rate of unbinding from inhibitory site & $b_i$ & varies \\
\hline
\rule{0pt}{4mm}
~Binding radius & $\varrho$ & $\SI{0.03}{\micro\meter}$ \\
\hline
\rule{0pt}{4mm}
~Unbinding radius & $\sigma$ & $\SI{0.015}{\micro\meter}$ \\
\hline
\rule{0pt}{4mm}
~BD time step & $\Delta t$ & $\SI{0.1}{\milli\second}$ \\
\hline
\rule{0pt}{4mm}
~Edge length of BD regime & $L_{BD}$ & $\SI{1}{\micro\meter}$ \\
\hline
\rule{0pt}{4mm}
~Compartment size & $h$ & $\SI{0.2}{\micro\meter}$ \\
\hline
\end{tabular}
\vskip 3mm
\caption{Parameter values for the spatial stochastic simulations, 
for physiologically relevant conditions~{\rm\cite{Rudiger2010}}.}
\label{tab:parameter}
\end{table}

\subsection{Diffusion - Brownian dynamics}

A versatile method for the simulation of particles in a solution is given by
Brownian Dynamics (BD). Collisions of particles with solution molecules lead to
overdamped dynamics and random forcing on a sufficiently long time
scale~\cite{Erban:2014:MDB}.  Assuming that there are $Q(t)$ free ions in the
simulation domain at time $t$, the equation for Brownian motion of ions is
given by
\begin{equation}
\label{eq:brownian-dynamics}
\text{d}\mathbf{X}_j=\sqrt{2D} \, \text{d}\mathbf{W}_j
\qquad j=1,2,\dots,Q(t)\;,
\end{equation}
where $D$ is the diffusion coefficient of ions in solution,
$\mathbf{X}_j(t) \in \Omega$ describes the trajectory of the $j$-th ion, and
$\mathbf{W}_j$ is a three-dimensional vector of independent Wiener
processes. This approach dramatically reduces the dimensionality of the problem
compared to molecular dynamics approaches wherein the degrees of freedom of
every participating molecule need to be taken into account. Nevertheless, the
computational load is still high compared to deterministic PDE-based approaches
to diffusion. There are a number of approaches for simulating
(\ref{eq:brownian-dynamics}) in the literature, ranging from discretization with
a fixed time step~\cite{Andrews:2004:SSC} to event-based
methods~\cite{vanZon:2005:GFR,Opplestrup:2009:FKM}.  In this paper, we
discretize time with the time step $\Delta t$ and use the Euler-Maruyama
discretization of equation~(\ref{eq:brownian-dynamics}), i.e. the position of
the $j$-th ion is updated according to
\begin{equation}
\label{eq:browndyn}
\mathbf{X}_j(t+\Delta t)
=
\mathbf{X}_j(t)
+
\sqrt{2D\Delta t}\,
\boldsymbol{\xi}_{j}\;,
\qquad j=1,2,\dots,Q(t)\;,
\end{equation}
where $\boldsymbol{\xi}_j$ is a vector of three independent normally 
distributed random numbers with zero mean and unit variance.

\subsection{Domain geometry and boundary conditions}
\label{secgeometry}

In our simulations, the computational domain is given as cube $\Omega=[0,L]^3$
where the value of $L$ is specified together with other parameters in
Table~\ref{tab:parameter}. \ca{} ions are able to undergo free diffusion in
$\Omega$ which we simulate using (\ref{eq:browndyn}). A cluster of nine IP$_3$R
channels is positioned in a $3\times3$ grid with grid constant
$\ell=\SI{0.15}{\micro\meter}$, centered in the $z=0$ plane, i.e.  the positions
of nine channels in the cluster are given as
\begin{equation}
\left[ \frac{L}{2},\frac{L}{2},0 \right], 
\quad 
\left[ \frac{L}{2} \pm \ell,\frac{L}{2},0 \right], 
\quad 
\left[ \frac{L}{2}, \frac{L}{2} \pm \ell,0 \right], 
\quad 
\left[ \frac{L}{2} \pm \ell, \frac{L}{2} \pm \ell,0 \right].
\label{channelpos}
\end{equation}
We found no significant effects from varying the channel spacing $\ell$, hence
we chose this particular cluster configuration for ease of implementation. Ions
bind to and dissociate from binding sites on the channels. The boundaries of the
computational domain at $x=0$, $x=L$, $y=0$, $y=L$, and $z=L$ are
constant-concentration boundaries, hence they absorb and introduce ions, such
that in the absence of open channels the concentration of \ca{} in the
computational domain is held at its equilibrium value, $c_0$, on average. The
boundary at $z=0$ is reflective, corresponding to the membrane of the ER. Hence,
ions can enter and leave the domain via the boundaries, in addition to being
introduced through open channels.

If the channels are closed, then the average number of free ions 
in the computational domain, $\langle Q(t) \rangle$, is
equal to $c_0 | \Omega |$ where $| \Omega | = L^3$ is the volume
of $\Omega$. Using our parameter values (see Table~\ref{tab:parameter}),
$c_0 | \Omega | \approx 1.5 \times 10^3$. The average number of ions
in the simulation domain, $\langle Q(t) \rangle$, increases
when the channels are open (by around two orders of magnitude). In order 
to simulate the system over sufficiently long time intervals, we will 
use a multiscale approach, which we describe in detail in
Section~\ref{sec:multiscale}.

\subsection{Stochastic channel binding model}

The conformational changes between the open and closed states 
of IP$_3$R channels are controlled by the binding of \ca{} to 
activating and inhibitory binding sites. These channels consist 
of $4$ subunits, that can be itself in an active or 
a neutral/inhibited state. Each subunit has three different binding
sites: An activating binding site for \ca{} ions, an inhibitory 
binding site for \ca{} ions, and an IP$_3$ binding site.

\begin{figure}[tbp]
\includegraphics[]{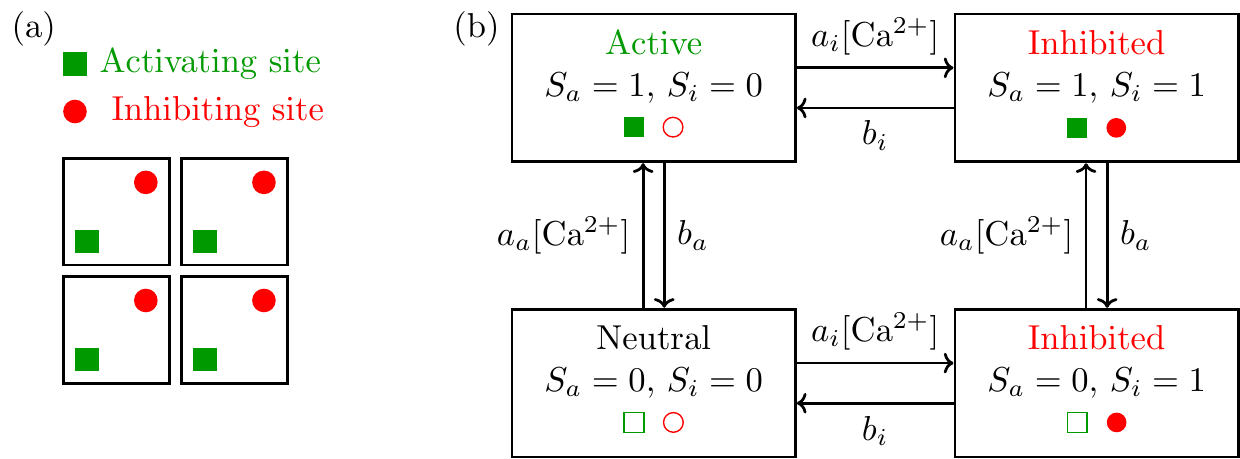}
\caption{{\rm (a)} 
{\it Sketch of a channel, containing $4$ subunits with an activating
ion binding site (green square) and an inhibiting site (red circle).}
{\rm (b)} 
{\it State space of a single subunit. Ions can bind to the activating 
and the inhibiting sites. A subunit is active only if it has an occupied
activating site as well as an empty inhibiting site.}}
\label{fig:stochchanmod}
\end{figure}

To accurately capture channel gating events, we employ a simplified
DeYoung-Keizer model~\cite{DeYoung1992}. Here, we disregard IP$_3$ 
dynamics and consider the effects of IP$_3$ only via their influence 
on the dissociation constant $b_i$ of \ca{} ions from inhibitory 
binding sites. Free \ca{} ions can bind to the activating and 
inhibitory sites, while bound ions can dissociate from occupied 
sites. Therefore, there are two reversible reactions in 
our model for each subunit of a channel:
\begin{equation}
\label{eq:reactions}
\begin{split}
\text{Ca}^{2+}
+
\big\{ S_a=0 \big\}
&
\; 
\mbox{\raise -0.9 mm \hbox{$
\displaystyle
\mathop{\stackrel{\displaystyle\longrightarrow}\longleftarrow}^{a_a}_{b_a}
$}}
\;
\big\{ S_a=1 \big\} \\
\text{Ca}^{2+}
+
\big\{ S_i = 0 \big\}
&
\; 
\mbox{\raise -0.9 mm \hbox{$
\displaystyle
\mathop{\stackrel{\displaystyle\longrightarrow}\longleftarrow}^{a_i}_{b_i}
$}}
\;
\big\{ S_i=1 \big\}
\end{split}
\end{equation}
The rates $a_a$ and $a_i$ describe the binding affinity of a \ca{} ion 
to an activating and inhibitory binding site, respectively, while 
the off-rates $b_a$ and $b_i$ describe the corresponding dissociation
reaction rates. The variables $S_a$ and $S_i$ describe the binding 
site state and can take values of $0$ and $1$. Channels consist 
of four subunits, with each subunit having one activating and one 
inhibitory \ca-binding site; see Figure~\ref{fig:stochchanmod}. 
A model subunit can then be in three distinct states: neutral (no 
binding site occupied), active (only the activating site
occupied) and inhibited (if the inhibitory site is occupied 
regardless of the state of the activating site). A channel then 
opens when at least three of its four subunits are in the active 
state.

\subsection{Ion binding dynamics}
\label{secBDbinding}

In order to precisely capture channel opening and closing dynamics, we need to
implement a reversible BD binding model of \ca{} ions to their corresponding
binding sites. Ions become binding candidates as soon as they enter a
half-sphere of radius $\varrho$ around a binding site (the channels are
positioned on the ER membrane, hence only the half-sphere above them is
available for ion binding). They are then allowed to bind to the site with a
probability $P_\lambda$ per time step that they spend in the binding
region~\cite{Lipkova2011,Erban2009}. An ion bound to an activating site
(resp. inhibitory site) is allowed to dissociate with a probability
$1-\exp(- b_a \, \Delta t)$ (resp. $1-\exp(- b_i \, \Delta t)$) and placed at a
distance of $\sigma$ (unbinding radius) from the binding site. We use the values
of binding and unbinding radii, $\varrho$ and $\sigma$, as given in
Table~\ref{tab:parameter}. Their values are chosen to be reasonable, such that
no overlaps occur between channels and that they are large enough such that the
BD simulation time step can be chosen reasonably large. Then we can
pre-calculate the remaining parameter $P_\lambda$ (binding probability) before
the start of simulations using the approach described in~\cite[Section
5]{Lipkova2011}.

\subsection{Channel opening and calcium flux}

Each channel has four subunits. Let $S_a^{j,k}(t)$
(resp. $S_i^{j,k}(t)$), $j=1,2,\dots,9$, $k=1,2,3,4$, be the state 
of the activating (resp. inhibiting) site of the $k$-th subunit
of the $j$-th channel in the cluster. Then the number of active
subunits of the $j$-th channel at time $t$ is
$$
N_a^j(t)
=
\sum_{k=1}^4 S_a^{j,k}(t) \Big( 1 - S_i^{j,k}(t) \Big). 
$$
If $N_a^j(t) \ge 3$, then the channel is considered to be in
an open state. When a channel is open, new ions are introduced 
with a rate of $\SI{3.12e5}{\second\tothe{-1}}$ (corresponding 
to channel current
$I_C = \SI{0.1}{\pico\ampere}$~\cite{Bruno2010,Smith2009,Vais2010}) 
at the channel site, simulating the flux of \ca{} ions out of 
an active channel. The positions of released ions then evolve 
according to equation~(\ref{eq:browndyn}). 

\section{Multiscale approach}
\label{sec:multiscale}

\begin{figure}[tb]
\centering
\includegraphics[]{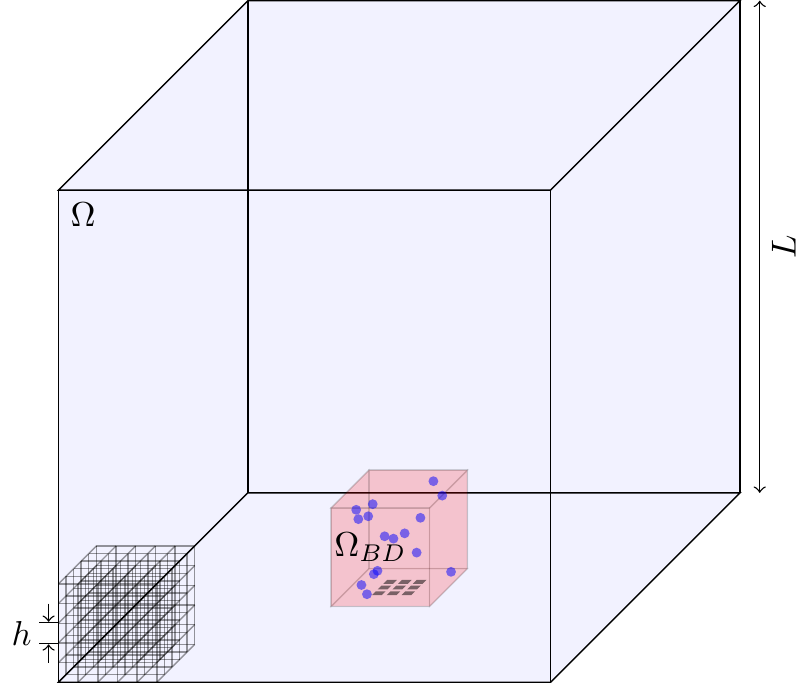}
\caption{{\it Sketch of the computational domain $\Omega$. It consists of two
    regions: The compartment-based part $\Omega \setminus \Omega_{BD}$
    (everywhere except for the red box in the bottom center, with the
    compartment size $h$ illustrated in the left bottom of the picture) and the
    BD domain $\Omega_{BD}$ (shown as a red box in the bottom center).  The
    boundary in both $x$- and $y$-directions as well as the boundary at $z=L$
    are constant-concentration boundaries, while the boundary at $z=0$ is
    reflective. The channel cluster is positioned at the center of the $z=0$
    boundary (indicated as black squares), see} $(\ref{channelpos})$.  }
\label{fig:modeldomain}
\end{figure}

During a puff event, a large number of ions are released into the 
cytoplasm. If all channels are open continuously, then one can
estimate, for the parameter values given in 
Table~\ref{tab:parameter}, that our computational domain 
may contain of the order of $10^5$ ions during the peak of a
puff. Since we are interested in time scales of minutes, tracking 
the individual position of this number of ions via Brownian 
dynamics becomes infeasible. However, high accuracy and individual 
ion positions are only needed in the vicinity of channel sites in order 
to ensure accurate implementation of BD binding dynamics, described
in Section~\ref{secBDbinding}. Therefore, we split our computational 
domain into two regions: a cube
$$
\Omega_{BD} 
= 
\left[ \frac{L - L_{BD}}{2}, \frac{L+ L_{BD}}{2} \right]
\times
\left[ \frac{L - L_{BD}}{2}, \frac{L+ L_{BD}}{2} \right]
\times
\left[ 0, L_{BD} \right]
$$
containing the channel sites (red region in Figure~\ref{fig:modeldomain}), 
as well as the remaining space $\Omega \setminus \Omega_{BD}$ in which 
we will use a coarser description of ion movement as described in the 
next subsection. Here, $L_{BD} < L$ is the length of the edge of
the cube $\Omega_{BD}$. We use $L_{BD} = L/5$ in our simulations
(see Table~\ref{tab:parameter}). 
In particular, we use BD simulations in a small fraction of 
$1/5^3 \approx 0.8 \%$ of the computational domain $\Omega$.

\subsection{Compartment-based model for diffusion}

We subdivide the region $\Omega \setminus \Omega_{BD}$ into compartments (cubes)
with size $h$ (illustrated in the bottom left of Figure~\ref{fig:modeldomain})
and employ a compartment-based method to simulate the movement of
ions~\cite{Erban:2007:PGS}.  Ions are allowed to move between adjacent
compartments with a rate of $d=D/h^2$. The compartment-based algorithm only
stores and evolves the number of ions in each compartment (rather than following
individual ions). An event-based stochastic simulation algorithm is used to
efficiently simulate this system. Several equivalent methods have been developed
in the literature, such as the Gillespie algorithm~\cite{Gillespie1976}, the
Next Reaction Method~\cite{Gibson2000}, the Next Subvolume
Method~\cite{Elf2004,Hattne2005}, as well as the Optimized Direct
Method~\cite{Cao2004}.

In this paper, we employ the Next Reaction Method~\cite{Gibson2000}. For each
possible move between two neighboring compartments, a putative time for the next
jump of an ion to occur is calculated. It is given by
\begin{equation}
t-\frac{\ln(r)}{d \, A_c(t)}
\label{jumptimes}
\end{equation}
where $t$ is the current simulation time, $d=D/h^2$ is the jump rate per one
ion, $A_c(t)$ is the current number of ions in the compartment from which an ion
is jumping and $r$ is a random number uniformly distributed in the interval
$(0,1)$.  Clearly, the putative jump time (\ref{jumptimes}) is infinity if
$A_c(t)=0$, i.e. if the corresponding compartment is empty. The putative times
are smaller (on average) if the corresponding compartment contains more ions.
 
The putative jump times are inserted into a priority queue (a heap data
structure~\cite{Cormen2009}), which enables us to efficiently extract the
earliest jump time and thus the next jump. This move is then performed, and the
numbers of ions in compartments (and the corresponding putative times and their
entries in the priority queue) are updated. We then iterate this process by
finding the minimal putative time and performing the corresponding ion jump at
each iteration.

At the boundaries of the computational domain $\Omega$ (except for 
the boundary at $z=0$, which is reflective), ions are absorbed 
(i.e. they leave the domain) or can enter with rates
consistent with a constant equilibrium concentration of $c_0$ outside the
domain. The jump rate from outside the domain into a compartment just 
inside the domain boundary is $c_0  D  h$ which is used,
instead of $d A_C(t)$, in (\ref{jumptimes})
to compute the corresponding putative times. 

\subsection{Coupling the BD simulation 
in $\Omega_{BD}$ with the compartment-based approach
in $\Omega \setminus \Omega_{BD}$}

Several methods exist in order to couple the BD and compart\-ment-based methods
across their interface, such as the two-regime method
(TRM)~\cite{Flegg2012,Flegg2014} or the ghost-cell
method~\cite{Flegg2015}. Here, we employ the TRM.  In order to accurately
capture diffusion across the interface, the jump rates from adjacent
compartments into $\Omega_{BD}$ need to be adjusted from the bulk rate $d=D/h^2$
to the interface rate~\cite{Flegg2012}
\begin{equation}
d_i=\frac{2}{h}\sqrt{\frac{D}{\pi \, \Delta t}}.
\label{trm1}
\end{equation}
These jump rates are used, instead of $d$, in (\ref{jumptimes})
to compute the corresponding putative times. If the chosen move in the compartment regime is a jump from a compartment adjacent to the 
interface into $\Omega_{BD}$, the occupancy number of the compartment 
is reduced by one and a new ion is added in $\Omega_{BD}$
at the distance $x$ from the interface which is sampled from
the probability distribution~\cite{Flegg2012}
\begin{equation}
f(x)=\sqrt{\frac{\pi}{4D\Delta t}}\,\text{erfc}
\Biggl[\frac{x}{\sqrt{4D\Delta t}}\Biggr],
\label{trm2}
\end{equation}
where $\text{erfc}$ is the complementary error function.

The rate (\ref{trm1}) and the distribution (\ref{trm2}) are used
to transfer ions from $\Omega \setminus \Omega_{BD}$ into
$\Omega_{BD}$. In the opposite direction, the TRM transfers
any ion from $\Omega_{BD}$ which during the time step interacts
with the interface. For a detailed discussion and the derivation 
of the probabilities and jump rates above, please see references~\cite{Flegg2012,Flegg2014,Flegg2015}. We employ 
the particle-based simulation library package \emph{Tyche} 
which implements the TRM~\cite{Tyche}. The TRM has also been
recently implemented in Smoldyn~\cite{Robinson}.

\section{Results}
\label{sec:results}

In the following section we present the results of our findings. 
Simulations were performed with the parameter values listed in
Table~\ref{tab:parameter} unless noted differently. In 
Section~\ref{secpstat}, we discuss \ca{} puff statistics from 
simulation runs. We follow this by investigating the regimes in 
which puffs are visible in Section~\ref{secinhtrans}. A simple 
mean-field theory is then presented in Section~\ref{secmeanfield}.
It describes the transition between puffs and perpetually open channels.

\subsection{Puff statistics}
\label{secpstat}

\begin{figure}[tbp]
\centering
\includegraphics[page=1]{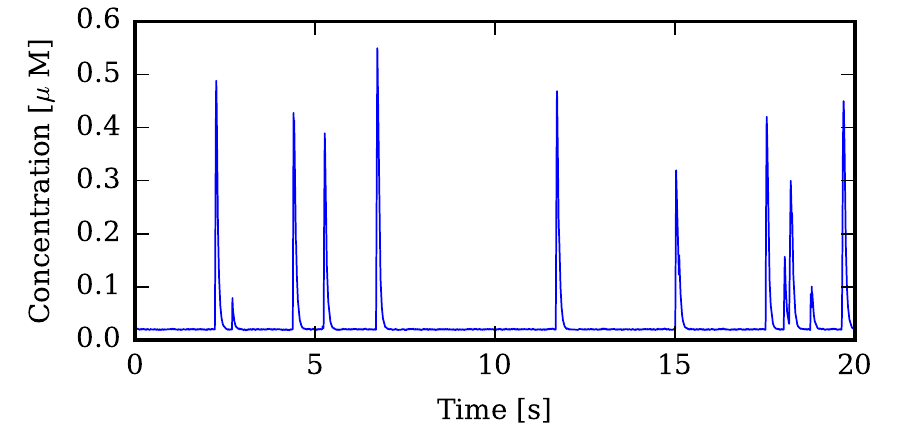}
\caption{Example concentration time trace for $a_i=\SI{1}{\brate}$ and
$b_i=\SI{1}{\ubrate}$.}
\label{fig:concentration-time}
\end{figure}

Figure~\ref{fig:concentration-time} displays a sample time trace of the \ca{}
concentration in the computational domain for $a_i=\SI{1}{\brate}$ and
$b_i=\SI{1}{\ubrate}$ for a time range of $\SI{20}{\second}$. The ion
concentrations shows erratic calcium puffs with an amplitude of up to
$\SI{0.5}{\micro\Molar}$. The constant background of
$c_0=\SI{0.02}{\micro\Molar}$ corresponds to approximately $12$ ions per
$\si{\micro\meter\tothe3}$ and justifies our use of Brownian dynamics because
fluctuations in such a small number of ions influence binding to the
channels. The puffs are characterized by a sharp, almost instant increase in
concentration, followed by an approximately exponential decay after the channel
cluster closes. Puffs are separated by a refractory phase in which channel
subunits are inhibited and the cluster cannot open.

\begin{figure}[tbp]
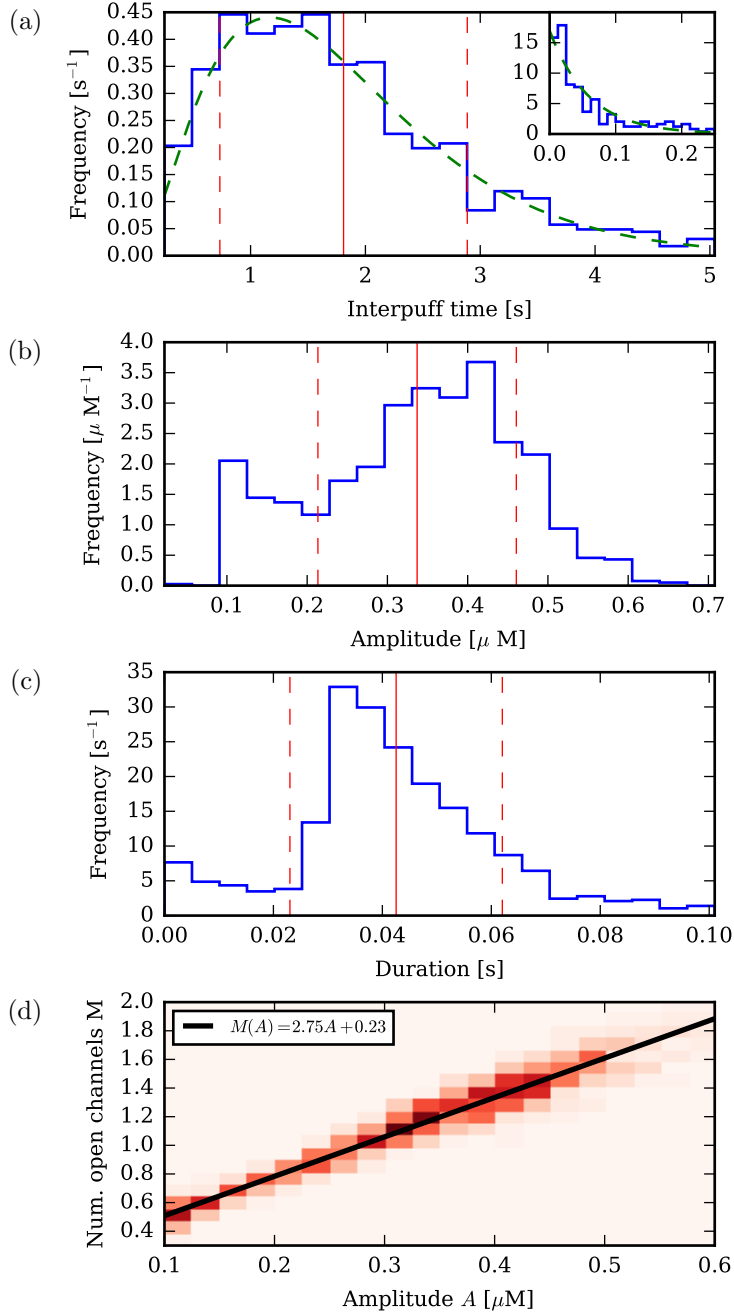

\centering
\begin{tikzpicture}
  \node[anchor=north west] (a) at (0,0) {\includegraphics[page=4,trim=0 0 0 5]{result_figures.pdf}};
  \node[anchor=north east] at (a.north west) {(a)};
  \node[anchor=north west] (b) at (a.south west) {\includegraphics[page=2,trim=0 0 0 5]{result_figures.pdf}};
  \node[anchor=north east] at (b.north west) {(b)};
  \node[anchor=north west] (c) at (b.south west) {\includegraphics[page=3,trim=0 0 0 5]{result_figures.pdf}};
  \node[anchor=north east] at (c.north west) {(c)};
  \node[anchor=north west] (d) at (c.south west) {\includegraphics[page=5,trim=0 0 0 5]{result_figures.pdf}};
  \node[anchor=north east] at (d.north west) {(d)};
\end{tikzpicture}
\vspace{-0.3cm}
\caption{Calcium puff statistics for a set of $N_p = \num{1151}$ puffs from
  simulations with a total duration of \num{3} simulation hours. The mean and
  standard deviations of the distributions are indicated by the solid and dashed
  red lines, respectively. {\rm (a)} Inter-puff time distribution for \num{545}
  puff intervals after threshold filtering $(\ref{thrfilt})$. The mean
  inter-puff time is $\num{1.81}\pm\SI{1.077}{\second}$. The green dashed line
  displays Gamma distribution $(\ref{eq:gamma-dist})$ with $\kappa=\num{2.82}$
  and $\theta=\SI{0.64}{\second}$. The inset shows the distribution of interpuff
  times smaller than $\SI{250}{\milli\second}$ with a mean of
  $\num{0.59}\pm\SI{0.061}{\second}$. {\rm (b)} Amplitude distribution. The mean
  puff amplitude is $\num{0.34}\pm\SI{0.12}{\micro\Molar}$ (indicated by the red
  vertical lines). {\rm (c)} Puff duration (full duration at half amplitude)
  distribution. The mean puff duration is
  $\num{0.04}\pm\SI{0.02}{\second}$. {\rm (d)} A two-dimensional histogram
  showing the strong correlation between the average number of open channels
  (over a puff's duration) with the puff amplitude. The black line indicates a
  linear fit to the correlation data.}
\label{fig:puff-statistics}
\end{figure}

We analyze the concentration data given as the time values $t_j$ and
concentration values $c_j$, $j=1,2,\dots,N$, (where $N$ is the index
of the last data point of the simulation run) by choosing a puff-starting threshold
concentration value 
\begin{equation}
c_t=\langle c_j \rangle +\sqrt{\text{Var}[c_j]},
\label{ctdef}
\end{equation}
i.e. one standard deviation above the mean concentration $\langle c_j
\rangle$.
For the data set shown below, the threshold was $c_t\approx 0.1\mu M$. The averaging in
(\ref{ctdef}) is taken over all values of $j$, $j=1,2,\dots,N$, i.e.  the sample
mean and variance in (\ref{ctdef}) are estimated by
\begin{equation}
\langle c_j \rangle
=
\frac{1}{N}
\sum_{j=1}^N
c_j
\qquad
\mbox{and}
\qquad
\text{Var}[c_j]
= 
\frac{1}{N-1}
\sum_{j=1}^N
\left( c_j - \langle c_j \rangle \right).
\label{averdef}
\end{equation}
Puffs are identified by the concentration crossing the threshold point
(\ref{ctdef}). Puff ending points are identified via a re-crossing of 
this value. The crossing points are given in the ordered index set
$$
\mathcal{S}=\{j\;|\;c_{j-1}<c_t\;\text{and}\;c_j\ge c_t,
\; j=2,3,\dots,N-1\}
$$ 
(puff starting indices) and
$$
\mathcal{E}=\{j\;|\;c_j>c_t\;\text{and}\;c_{j+1}\le c_t,
\; j=2,3,\dots,N-1\}
$$
(puff ending indices). The sets $\mathcal{S}$ and $\mathcal{E}$ are 
enumerated by the puff index $1 \le k \le N_p$, where $N_p=|\mathcal{E}|$ 
is the number of finished puffs in the data set, i.e. we disregard 
the last index in $\mathcal{S}$ if the last puff did not finish
and denote
$$
\mathcal{S} = \{ s_1, s_2, \dots, s_{N_p} \}
\qquad \mbox{and} \qquad
\mathcal{E} = \{ e_1, e_2, \dots, e_{N_p} \},
$$
where the elements in $\mathcal{S}$ (resp. $\mathcal{E}$) are ordered,
i.e. $s_1 < s_2 < \dots < s_{N_p}$ 
(resp. $e_1 < e_2 < \dots < e_{N_p}$).
The inter-puff times are then given by
\begin{equation}
\mathcal{T}
=
\Big\{
t_{s_{k+1}}-t_{e_{k}} \, \big| \, 
s_{k+1} \in \mathcal{S}, \;
e_k \in \mathcal{E}, \;
1 \le k \le N_p-1
\Big\} 
\; \medcap \; (t_p,\infty),
\label{thrfilt}
\end{equation}
where the threshold $t_p = 0.25$ s. This threshold filtering decreases the
number of puffs considered by removing the puffs which are not well
separated. Indeed, the inset in Figure~\ref{fig:puff-statistics}(a) shows that
the distribution of interpuff times below the threshold follows an exponential
distribution and is due to random channel reopenings stemming from residual
calcium. The distribution of inter-puff times (histogram of set $\mathcal{T}$)
is shown in Figure~\ref{fig:puff-statistics}(a) for a set of \num{954} puff
intervals. Its shape is similar to a Gamma distribution
\begin{equation}
\label{eq:gamma-dist}
p(\tau) =\frac{1}{\Gamma(\kappa)\theta^\kappa}\tau^{\kappa-1}e^{-\tau/\theta}\;.
\end{equation}
Indeed, when choosing $\kappa=\num{2.82}$ and $\theta=\SI{0.64}{\second}$ in
equation~(\ref{eq:gamma-dist}) (such that the resulting distribution has the
same mean and variance as our data), the comparison is good.  As the Erlang
distribution (the special case of a Gamma distribution with
$\kappa\in\mathbb{N}$) results from summing $\kappa$ exponentially-distributed
random variables, we can speculate that an effective model of calcium channel
opening might be described by a three-stage binding/unbinding process. The
maximum of the inter-puff time distribution, $\tau_{m}\approx\SI{1.81}{\second}$,
can be interpreted as the typical time between puffs in this system.

The distribution of puff amplitudes are determined by taking the 
maximum value in a previously-determined puff interval
$$
\mathcal{A}
=
\left\{ a_k \, \big| \, a_k = \max_{s_k \le j \le e_k} c_j,
\; \mbox{where} \; 
s_k \in \mathcal{S}, \;
e_k \in \mathcal{E}, \;
1 \le k \le N_p \;
\right\}.
$$
The amplitude distribution (histogram of set $\mathcal{A}$) is shown 
in Figure~\ref{fig:puff-statistics}(b) and mirrors the broad distribution
characterized in experiments~\cite{Dickinson2013}.

The puff durations are calculated by considering the full duration at half
maximum, which is a common criterion for determining the puff duration in the
literature~\cite{Dickinson2013}. To this end, we calculate 
the indices of crossings of the half maximum threshold on the rising 
slope of a puff
\begin{eqnarray*}
\mathcal{R}
=
\Big \{
r_k
\, &\big|& \, 
r_k = \max \{j \, | \, c_{j-1} < a_k/2 \;\text{and}\;
c_{j}\ge a_k/2\;
\text{and}\; s_k \le j \le e_k\}, \\
&&
\; \mbox{where} \; 
s_k \in \mathcal{S}, \;
e_k \in \mathcal{E}, \;
a_k \in \mathcal{A}, \;
1 \le k \le N_p \;
\Big\}\;,
\end{eqnarray*}
and the falling slope of a puff
\begin{eqnarray*}
\mathcal{F}
=
\Big \{
f_k
\, &\big|& \, 
f_k = \min \{j \, | \, c_{j} > a_k/2 \;\text{and}\;
c_{j+1}\le a_k/2\;
\text{and}\; s_k \le j \le e_k\}, \\
&&
\; \mbox{where} \; 
s_k \in \mathcal{S}, \;
e_k \in \mathcal{E}, \;
a_k \in \mathcal{A}, \;
1 \le k \le N_p \;
\Big\}\;.
\end{eqnarray*}
The puff durations are then calculated via
$$
\mathcal{D}
=
\big\{
t_{f_{k}}-t_{r_{k}} \, | \, 
f_{k} \in \mathcal{F}, \;
r_k \in \mathcal{R}, \;
1 \le k \le N_p
\big\}.
$$
The distribution of puff durations (histogram of set $\mathcal{D}$) 
is plotted in Figure~\ref{fig:puff-statistics}(c). The distribution is 
sharply peaked around $\SI{0.03}{\second}$ and has a long tail stemming 
from repeated cluster re-openings during the concentration decay phase. 
The mean puff lifetime is within the range 40--70 ms measured in
experiments~\cite{Dickinson2013}.

Figure~\ref{fig:puff-statistics}(d) displays the correlation between the number
of open channels averaged over a puff's duration $M$ and the puff amplitude
$A$. By finding the average $M(A)$ for each value of $A$, and performing a
linear fit, we find the linear relationship $M(A)=2.75A+0.23$.

\subsection{Inhibition transition}
\label{secinhtrans}

\begin{figure}[p]
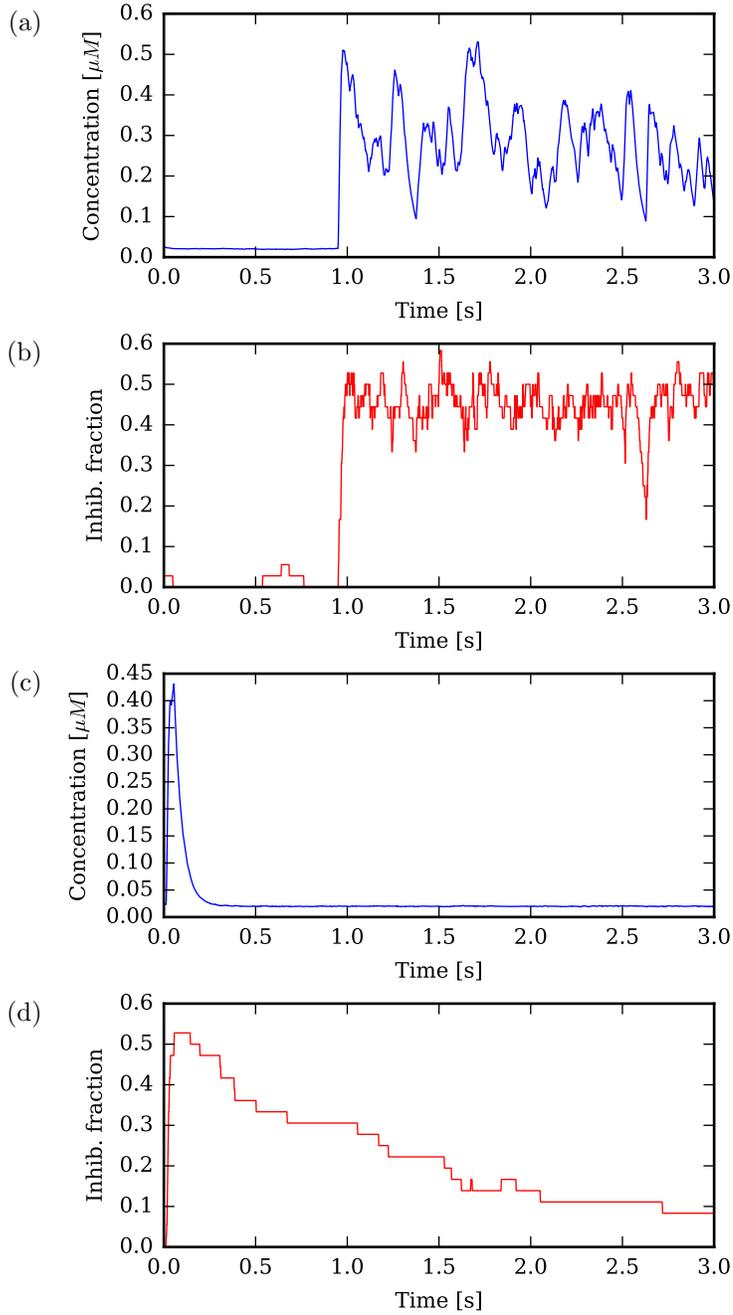

\centering
\begin{tikzpicture}
  \node[anchor=north west] (a) at (0,0) {\includegraphics[page=9,trim=0 0 0 5]{result_figures.pdf}};
  \node[anchor=north east] at (a.north west) {(a)};
  \node[anchor=north west] (b) at (a.south west) {\includegraphics[page=10,trim=0 0 0 5]{result_figures.pdf}};
  \node[anchor=north east] at (b.north west) {(b)};
  \node[anchor=north west] (c) at (b.south west) {\includegraphics[page=11,trim=0 0 0 5]{result_figures.pdf}};
  \node[anchor=north east] at (c.north west) {(c)};
  \node[anchor=north west] (d) at (c.south west) {\includegraphics[page=12,trim=0 0 0 5]{result_figures.pdf}};
  \node[anchor=north east] at (d.north west) {(d)};
\end{tikzpicture}
\caption{{\rm (a)} \it Calcium concentration and {\rm (b)} fraction of inhibited
  channel subunits as a function of time for $a_i=\SI{0.1}{\brate}$ and
  $b_i=\SI{5}{\ubrate}$.  Other parameter values are given in
  Table~$\ref{tab:parameter}$.  Note that the channel cluster stays perpetually
  active in this configuration.
  {\rm (c)} \ca{} concentration and {\rm (d)} fraction of occupied inhibitory
  binding sites in the channel cluster as a function of time for the inhibitory
  site binding parameters $a_i=\SI{0.1}{\brate}$ and
  $b_i=\SI{0.1}{\ubrate}$. Other parameter values are given in
  Table~$\ref{tab:parameter}$.  Note that the time scale of inhibition decay is
  long enough for the calcium concentration to decay completely, thus the
  channel cluster closes and we can identify well-defined puffs.}
\label{fig:conc-inh-frac-time}
\end{figure}

Figure~\ref{fig:conc-inh-frac-time} shows the results of a simulation run with
two different values of inhibitory site dissociation rate $b_i$, illustrating
two different modes of behavior. We use the same value of $a_i$ in all four
panels (the rest of parameter values are given in
Table~\ref{tab:parameter}). Figure~\ref{fig:conc-inh-frac-time}(a-b) show the
results of a simulation run with $b_i=\SI{5}{\ubrate}$ (corresponding to a
dissociation constant of $K_D=b_i/a_i=\SI{50}{\micro\Molar}$).  These simulation
parameters are consistent with the modeling of calcium puffs in the literature
and used in hybrid PDE-based models~\cite{Rudiger2010}.
Figure~\ref{fig:conc-inh-frac-time}(a) shows the \ca{} concentration in the
computational domain $\Omega$, while Figure~\ref{fig:conc-inh-frac-time}(b)
displays the fraction of occupied inhibitory sites in the channel cluster as a
function of time. Due to the large dissociation rate, the necessary level of
bound inhibitory sites can never be sustained long enough for all the channels
in the cluster to close at the same time. Therefore, the \ca{} concentration is
kept at a level where empty activating sites are immediately filled and the
channel cluster stays perpetually open. Hence, puff termination in this case can
be speculated to be facilitated by a mechanism other than channel inhibition,
such as ER \ca{} reservoir depletion or a mechanism involving dissociation of
IP$_3$~\cite{rudiger2012termination}.  We checked the influence of lowering the
binding radius $\varrho$ to $\SI{6}{\nano\meter}$ and found that it had no
effect on the channel cluster closing.

In contrast, a lower inhibitory site dissociation constant yields well-defined
calcium puffs in our particle-based simulation scheme.
Figure~\ref{fig:conc-inh-frac-time}(c-d) display data from simulation runs with
$b_i=\SI{0.1}{\ubrate}$ (which corresponds to
$K_D=b_i/a_i=\SI{1}{\micro\Molar}$).  Here, inhibitory sites binding is
sustained on a high level for a long enough time such that all channels close
and the excess \ca{} is removed.  Hence there are time intervals when the
cluster concentration decreases and reaches the equilibrium $Ca^{2+}$
concentration $c_0$.  Puffs are therefore clearly delineated and separated with
a well-defined inter-puff time.

Hence there exist two regimes: a puff regime
[Figures~\ref{fig:conc-inh-frac-time}(c) and~\ref{fig:conc-inh-frac-time}(d)]
and a regime with perpetually open channel clusters
[Figures~\ref{fig:conc-inh-frac-time}(a) and~\ref{fig:conc-inh-frac-time}(b)].
We now proceed to characterize concentration and open channel time traces to
find under which conditions well-defined puffs are possible. To this end, we use
the ``puff score'' characterization function introduced in~\cite{Hao2009}, which
quantifies the spike-ness of a given time trace of the number of open
channels. We denote the number of open channels at time point $t_j$ by $O_j$,
$j=1,2,\dots,N$, i.e.  $O_j \in \{ 0, 1, 2, \dots, C \}$ where $C=9$ in our
simulations. Then the puff score is defined by
\begin{equation}
\label{eq:puff-score}
\mbox{[PS]}(a_i,b_i)
=
\frac{1}{C}\frac{\text{Var}[O_j]}{\langle O_j \rangle}\;.
\end{equation}
where the averages are again take over all values of $j$, $j=1,2,\dots,N$
(compare with (\ref{averdef})). The puff score (\ref{eq:puff-score})
can take values in  $[\num{0},\num{1}]$. A 
puff score greater than $0.25$ indicates channel excitability and therefore 
the existence of puffs in the system.

\begin{figure}[t]
\centering
\includegraphics[page=6]{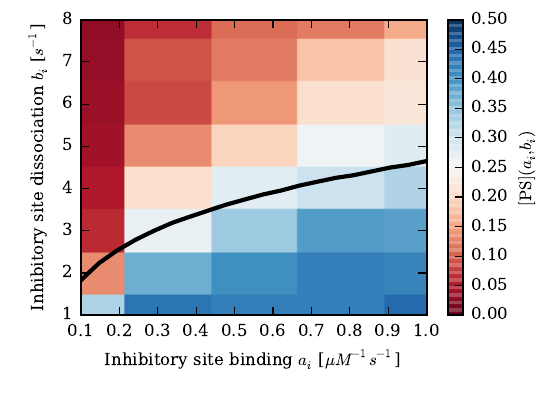}
\caption{{\it Phase plot of $a_i$ and $b_i$. Color indicates the puff
score $(\ref{eq:puff-score})$ for the given set of parameters listed
in Table~\ref{tab:parameter}. The black line shows the
numerically-determined phase boundary from the mean-field model
$\eqref{eq:mean-field-model-1}$--$\eqref{eq:mean-field-model-3}$.}}
\label{fig:phase-boundary}
\end{figure}

\begin{figure}[t]
\centering
\includegraphics[page=8]{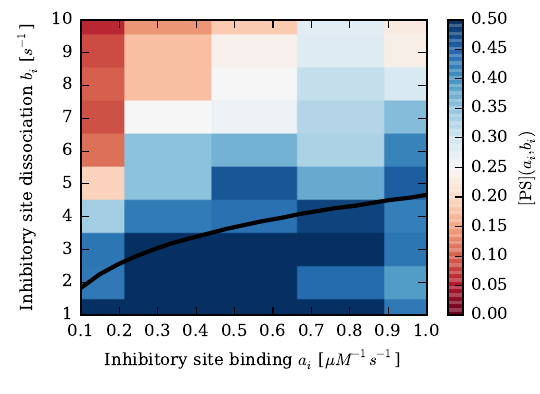}
\caption{{\it Phase plot of $a_i$ and $b_i$ for the hybrid scheme described in
    the text. Color indicates the puff score $(\ref{eq:puff-score})$ for the
    given set of parameters listed in Table~\ref{tab:parameter}. For comparison,
    the black line shows the numerically-determined phase boundary from the
    mean-field model
    $\eqref{eq:mean-field-model-1}$--$\eqref{eq:mean-field-model-3}$ for
    Figure~{\rm\ref{fig:phase-boundary}}.}}
\label{fig:regimes-hybrid}
\end{figure}

In order to visualize the two parameter regimes, we performed simulations for 
different inhibitory site binding parameters $a_i\in[0.1,1]\,\si{\brate}$ and
$b_i\in[1,10]\,\si{\ubrate}$ for a simulation time duration of
$\SI{100}{\second}$. We extracted the number of open channels over time 
and calculated the puff score (\ref{eq:puff-score}). 
Figure~\ref{fig:phase-boundary}
displays this quantity. The color indicates the value of $\mbox{[PS]}(a_i,b_i)$. 
The transition between the two regimes is not sharp, but gradual, especially 
for higher $a_i$. This is due to prolonged channel re-openings becoming more 
likely due to faster dissociation of bound inhibiting ions when $b_i$ approaches 
the transition. The phase boundary is consistent with a dissociation constant of
$\SI{4}{\micro\Molar}<K_D<\SI{10}{\micro\Molar}$.

To directly compare our results with previously-used schemes from the
literature, we devised a simplified hybrid scheme: The main difference between
our method and hybrid simulation algorithms is the much lower calcium
concentration in the vicinity of open channels and the resulting weaker channel
inhibition. Therefore, in the simplified hybrid scheme, whenever a channel is
open, we do not use the particle-based binding described in
section~\ref{secBDbinding}. Instead we assume a constant high calcium
concentration of $c_B=\SI{150}{\micro\Molar}$ (consistent with results from
hybrid simulations~\cite{Rudiger2010}) to generate random binding events to
inhibitory sites with a rate of $a_i c_B$, irrespective of any ions in the
vicinity. In all other respects, the simulation proceeds as described
previously. Figure~\ref{fig:regimes-hybrid} shows the resulting map of puff
scores.

\begin{figure}[t]
\centering
\includegraphics[page=7]{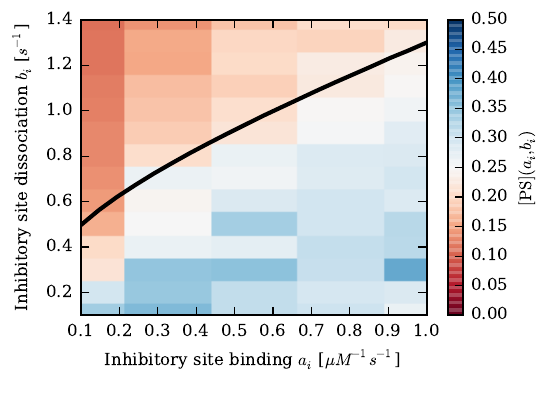}
\caption{{\it Phase plot of $a_i$ and $b_i$ with a low diffusion constant
    $D=\SI{20}{\micro\meter\tothe{2}\per\second}$. The other parameters are
    given in Table~\ref{tab:parameter}. Color indicates the puff score
    $(\ref{eq:puff-score})$. The black line shows the numerically-determined
    phase boundary from the mean-field model
    $\eqref{eq:mean-field-model-1}$--$\eqref{eq:mean-field-model-3}$. Note that
    the scale of the y-axis here is different from
    Figures~\ref{fig:phase-boundary} and~\ref{fig:regimes-hybrid}.}}
\label{fig:phase-boundary-d20}
\end{figure}

In order to study the influence of buffers (whose main effect is to slow down
ion diffusion \cite{keenersneyd}) on the boundary of the puff regime, we
performed a similar set of simulations with a lower \ca{} diffusion constant of
$D=\SI{20}{\micro\meter\tothe{2}\per\second}$. Because ions bound to buffer
molecules can be viewed as slowly diffusing ions, lowering the diffusion
constant is a way to model the effects of buffers~\cite{Wagner1994}. The result
is shown in Figure~\ref{fig:phase-boundary-d20}. With a lower diffusion
constant, the calcium concentration in the channel cluster nanodomains decays
more slowly. For puffs to exist, the time scale of the decay of inhibitory site
binding needs to be longer than the time scale of \ca{} decay. Hence, the
boundary separating the two regimes is pushed to smaller values of the
inhibitory dissociation rate $b_i$, corresponding to approximately
$\SI{1}{\micro\Molar}<K_D<\SI{2}{\micro\Molar}$. Hence, the effective diffusion
constant plays an important role in determining the boundary between the two
regimes. Note, that the overall variation of the puff score
$\mbox{[PS]}(a_i,b_i)$ is smaller compared to the case of
$D=\SI{220}{\micro\meter\tothe{2}\per\second}$, therefore \ca{} puffs become
less pronounced with slower ion diffusion. \ca{} buffers lead to additional
extrinsic noise due to binding and unbinding of calcium ions to buffer molecules
that enhances \ca{} fluctuations which might have effects on the puff statistics
on top of what we presented here~\cite{weinberg2014influence}.

\subsection{Mean-field model}
\label{secmeanfield}

In order to find an approximate phase diagram to determine the parameter regimes
in which calcium puffs occur, we develop a simplified non-spatial model. To this
end, we consider an ensemble of identical channels that interact by a shared
calcium domain (all-to-all coupling). The concentration $c(t)$ in the
$\SI{1}{\micro\meter\tothe{3}}$ cube around a channel undergoes exponential
decay with a phenomenological decay parameter $\lambda$ due to diffusive
equilibration, and a linear increase with an ion influx rate $\nu$ when the
channel is open. The open states of the ensemble of channels is determined by
the mean number of occupied activating and inhibitory binding sites per channel,
$a(t)$ and $b(t)$, in a similar way as in the spatial model above.  In a simple
representation of the subunit dynamics and their binding cooperativity we
require that a channel is open at a given time $t$ if $a(t)\ge 3$ and
$b(t)<2$. Here the variables $a(t)$ and $b(t)$ describe how many subunits, on
average, have activating and inhibitory ions bound to their respective binding
sites. They evolve according to the mass-action rate equations corresponding to
the reactions (\ref{eq:reactions}). Hence the model equations are:
\begin{align}
\label{eq:mean-field-model-1}
\frac{d c}{d t}&=\Theta(a-3) \, \Theta(2-b) \, \nu 
- \lambda \, (c-c_0) \;,\\
\label{eq:mean-field-model-2}
\frac{d a}{d t}&=a_a \, c \, (4-a) - b_a \, a\;,\\
\label{eq:mean-field-model-3}
\frac{d b}{d t}&=a_i \, c \, (4-b) - b_i \, b\;.
\end{align}
Here, $\Theta(x)$ is the Heaviside function with the properties
\[\Theta(x)=
\begin{cases}
  0 & \text{for}\,x<0 \\
  1/2 & \text{for}\,x=0 \\
  1 & \text{for}\,x>0 \;.
\end{cases}
\]
The first term on the right hand side of equation~(\ref{eq:mean-field-model-1})
describes the above-mentioned channel openings: The channels only open if three
subunits are active and not inhibited. The influx rate is determined via the
channel current $\nu=(2 e V)^{-1}I_C=\SI{518.28}{\micro\Molar\per\second}$ (where
$e=\SI{1.602e-19}{\coulomb}$ is the electron charge; the in-flowing ions are
assumed to be spread over a volume of $V=\SI{1}{\micro\meter\tothe{3}}$). This
value is also consistent with influx rates extracted from the rising flanks of
puffs in our simulations. The exponential decay parameter $\lambda$ was
determined by fitting an exponential decay to calcium puff simulation data. The
parameters of the mean-field model are summarized in
Table~\ref{tab:mean-field-parameters}.

\begin{table}[t]
\centering
\begin{tabular}{|l|c|l|}
\hline
\rule{0pt}{4mm}
$c_0$ & $\SI{0.02}{\micro\Molar}$ & Background Ca$^{2+}$ \\
\hline
\rule{0pt}{4mm}
$\nu$ & $\SI{5.18e2}{\micro\Molar\per\second}$ & Source rate \\
\hline
\rule{0pt}{4mm}
$\lambda$ & \begin{tabular}{@{}c@{}}$\SI{22.9}{\per\second}$ 
\rule{0pt}{4mm}
($D=\SI{220}{\micro\meter\tothe{2}\per\second}$) \\ $\SI{2.2}{\per\second}$ 
\rule{0pt}{4mm}
($D=\SI{20}{\micro\meter\tothe{2}\per\second}$)\end{tabular} & Ca$^{2+}$ decay \\
\hline
\rule{0pt}{4mm}
$a_a$ & $\SI{100}{\brate}$ & Rate of activating site binding \\
\hline
\rule{0pt}{4mm}
$b_a$ & $\SI{20}{\ubrate}$ & Rate of activating site unbinding \\
\hline
\rule{0pt}{4mm}
$a_i$ & $[0.1,1]\,\si{\brate}$ & Rate of inhibiting site binding \\
\hline
\rule{0pt}{4mm}
$b_i$ & $[0.1,8]\,\si{\ubrate}$ & Rate of inhibiting site unbinding \\
\hline
\end{tabular}
\vskip 3mm
\caption{Parameter values for the ODE mean-field model
\eqref{eq:mean-field-model-1}-\eqref{eq:mean-field-model-3}.}
\label{tab:mean-field-parameters}
\end{table}

\begin{figure}[tbp]
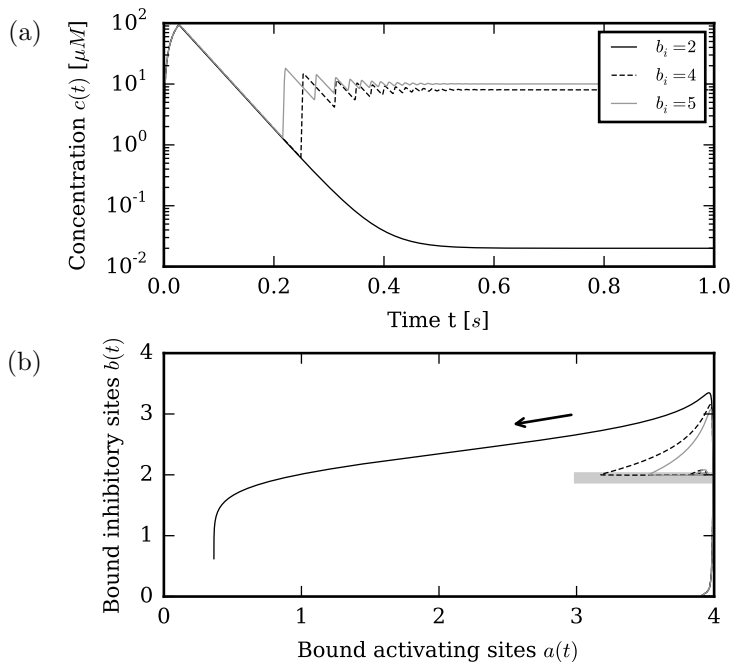

\centering
\begin{tikzpicture}
  \node[anchor=north west] (a) at (0,0) {\includegraphics[page=13,trim=0 0 0 5]{result_figures.pdf}};
  \node[anchor=north east] at (a.north west) {(a)};
  \node[anchor=north west] (b) at (a.south west) {\includegraphics[page=14,trim=0 0 0 5]{result_figures.pdf}};
  \node[anchor=north east] at (b.north west) {(b)};
\end{tikzpicture}
\caption{
{\rm (a)} 
{\it Concentration over time and} 
{\rm (b)}
{\it phase space trajectories of the mean field system
\eqref{eq:mean-field-model-1}-\eqref{eq:mean-field-model-3}. The
ratchet-like oscillation visible in the concentration traces for
$b_i=$\SIlist{4;5}{\ubrate} stems from the channels being
re-activated repeatedly due to insufficient inhibition.}}
\label{fig:ode-calc-inhibition}
\end{figure}

Figure~\ref{fig:ode-calc-inhibition} shows data from three representative
numerical solutions of the ODE
system~\eqref{eq:mean-field-model-1}-\eqref{eq:mean-field-model-3} where the
inhibitory binding rate is set to $a_i=\SI{0.5}{\brate}$. The initial conditions
are $c(0)=c_0$, $a(0)=4$ and $b(0)=0$. The concentration $c(t)$ in
Figure~\ref{fig:ode-calc-inhibition}(a) shows a puff with its characteristic
exponential decay to $c_0$. For $b_i=\SI{2}{\ubrate}$, $c(t)$ returns to its
equilibrium value $c_0$, while for higher values, it oscillates at a high
concentration level. These two states correspond to the ``puff'' regime and the
``always open'' regime, respectively. Figure~\ref{fig:ode-calc-inhibition}(b)
shows the corresponding trajectories in $a(t)$-$b(t)$ space. When the
trajectory, after its initial transient phase, does not touch the line segment
\mbox{$\{(a,2)$ for $a \in (3,4)\}$}, highlighted in gray in
Figure~\ref{fig:ode-calc-inhibition}(b), the concentration will decay to $c_0$
and the system returns to its original state. If, however the $a(t)$-$b(t)$
trajectory hits this boundary, the system stays excited and the channels stay
perpetually open. This can be translated into a temporal criterion by viewing
the time evolution of the system as a two-step process: (1) Find the time $\tau$
such that $b(\tau)=2$ (i.e. the time until the second Heaviside function in
equation~(\ref{eq:mean-field-model-1}) becomes zero); (2) find $\tau_1>\tau$ such
that $b(\tau_1)=2$ and $\tau_2>\tau$ such that $a(\tau_2)=3$. If
$\tau_2<\tau_1$, well-defined puffs are possible, otherwise $c(t)$ stays
elevated and channels stay perpetually open.

We now proceed to find the mean-field phase boundary between the ``puff'' regime
and the ``perpetually-open'' regime. Steady-state analysis of
equations~(\ref{eq:mean-field-model-1})-(\ref{eq:mean-field-model-3}) does not
yield the correct results in the latter regime due to the discontinuous nature
of the Heaviside functions in equation~(\ref{eq:mean-field-model-1}) and the
resulting temporal criterion above. The system rapidly switches between the
inhibited state with $b(t)>2$ and the non-inhibited state with $b(t)<2$, leading
to the observed decaying oscillations of $c(t)$ in
Figure~\ref{fig:ode-calc-inhibition}(a). Hence, we numerically integrate
equations~(\ref{eq:mean-field-model-1})-(\ref{eq:mean-field-model-3}) to test if
a given parameter set $a_i$, $b_i$ falls into one of the two regimes. We let the
system first evolve until $b>2$ and continue until either $a(t)$ becomes smaller
than $3$ or $b(t)$ becomes smaller than $2$. We then identify the regime the
system is in according to the criterion described above. A bisection algorithm
is used to find the boundary $b_i(a_i)$ between the two regimes. The black lines
in Figures~\ref{fig:phase-boundary},~\ref{fig:regimes-hybrid}
and~\ref{fig:phase-boundary-d20} show the extracted phase boundaries for the
parameter values given in Table~\ref{tab:mean-field-parameters}.

\section{Conclusions}
\label{sec:conclusions}

In this paper, we have reported a novel application of a particle-based spatial
algorithm for diffusion to investigate the influence of diffusive noise on the
dynamics of intracellular calcium release. Particle number noise in calcium
micro-domains has attracted interest in recent
studies~\cite{Flegg2013,weinberg2014influence,wieder2015exact,Ji2015}, also for L-type
and RyR calcium channels~\cite{Koh2006,Tanskanen2007,Hake2008}, and it is
important to clarify whether calcium diffusion as an additional noise source
needs to be incorporated to obtain a better understanding of sub-cellular
calcium signals.

In order to make this study feasible, we split the domain into a
compartment-based regime and a Brownian dynamics regime, coupled via the
TRM. This allowed us to model the dynamics of full puffs, including release of a
realistic number of ions and inhibition dynamics.  We extracted concentration
time traces and analyzed the resulting puff statistics. The inter-puff time
distribution as well as the distributions of puff amplitudes and lifetimes agree
qualitatively with experimental data in the literature~\cite{Fraiman2006}.

We then proceeded to analyze the binding parameter regimes under which
well-defined \ca{} puffs are possible. We found that, surprisingly, an
inhibitory binding site dissociation constant $K_D=\SI{50}{\micro\Molar}$
consistent with the literature and patch-clamp
experiments~\cite{shuai2009models}, does not yield puffs in our model. In this
parameter regime, channels stay perpetually open. In order to investigate the
transition between well-defined puffs and perpetually-open channels, we
characterized calcium concentration traces for various combinations of the
inhibitory site binding parameters. The phase boundary visible in our data is
consistent with a dissociation constant in the region
$\SI{4}{\micro\Molar}<K_D<\SI{10}{\micro\Molar}$. Lower values of the \ca{}
diffusion constant yield a phase boundary at smaller inhibitory site
dissociation rates and thus an even smaller dissociation constant. Given the
reliable puff generation and termination in previous studies based on fitted
gating models with large $K_D$~\cite{Rudiger2010,ullah2012multi} this is an
unexpected result.

Why does our model not show the same robust termination at large $K_D$ as the
hybrid approaches~\cite{rudiger2010calcium,Rudiger2010,ruediger2014stochastic}?
In the latter models the binding and unbinding to the receptors is stochastic
but the calcium distribution is calculated from deterministic reaction-diffusion
equations.  To test for the differences of puffs in both models we have
performed a set of simulations using large local \ca{} concentrations for open
channels, similar to what is obtained in the hybrid method. These simulations
confirm the robust termination in the hybrid scheme even at large $K_D$. In a
hybrid approach, a deep inhibitory state is achieved owing to the large local
nano-domain around each open channel displayed in the solution of deterministic
calcium equations~\cite{rudiger2010calcium}.  Thus, in the hybrid model, a
channel is not inhibited by shared calcium in the domain but by its own released
calcium. This effect has been termed self-inhibition \cite{ullah2012multi}.

In our simulations at large $K_D$, however, there is insufficient inhibition to
the channels during the early phase of a puff. This means that no or perhaps
only one subunit per channel binds inhibitory calcium, while reliable inhibition
requires binding of three or four calcium ions.  Our results suggest that
diffusive noise mixes calcium in the cluster domain and diminishes the localized
domains around open channels and self-inhibition.  Thus we are led to a model of
\ca{} puffs that is very different from the previous hybrid model. Our diffusive
model allows inhibition only from a much less localized concentration profile
and can therefore only be achieved at a much smaller dissociation constant of
inhibitory binding sites. This conclusion is also supported by the the
mean-field ODE model that we have devised and that captures the average binding
state of the cluster's activating and inhibitory sites as well as the resulting
\ca{} concentration in a shared and well-mixed micro-domain. This model displays
a sharp phase boundary between the two regimes, which agrees well with the data
from our spatial simulations.

Evidence for an inhomogeneous or homogeneous calcium distribution in the cluster
is hard to obtain directly from our simulations because of the short lifetime of
nano domains. In any case, our study highlights the role of the local calcium
concentration in the termination of puffs and shows that puffs are very
sensitive to fluctuations of residual calcium remaining after channel
closing. It has to be noted though, that, apart from the diffusive noise, there
are other differences of the current BD setup and the former hybrid
approaches. Notably, these include the presence of calcium-binding buffers and
the action of SERCA pump terms on the ER membrane boundary~\cite{Rudiger2010}
and it remains to be analyzed to which extent these differences affect puff
termination.

\bibliographystyle{siam}

\begin{thebibliography}{10}

\bibitem{Allbritton1992}
{\sc N.~L. Allbritton, T.~Meyer, and L.~Stryer}, {\em {Range of messenger
  action of calcium ion and inositol 1,4,5-trisphosphate.}}, Science (New York,
  N.Y.), 258 (1992), pp.~1812--1815.

\bibitem{Anderson2014}
{\sc M.~E. Anderson}, {\em {Three ways to die suddenly: Do they all require
  calcium calmodulin-dependent protein kinase II?}}, Transactions of the
  American Clinical and Climatological Association, 125 (2014), pp.~173--85.

\bibitem{Anderson2011}
{\sc M.~E. Anderson, J.~H. Brown, and D.~M. Bers}, {\em {CaMKII in myocardial
  hypertrophy and heart failure.}}, Journal of molecular and cellular
  cardiology, 51 (2011), pp.~468--73.

\bibitem{Andrews:2004:SSC}
{\sc S.~Andrews and D.~Bray}, {\em Stochastic simulation of chemical reactions
  with spatial resolution and single molecule detail}, Physical Biology, 1
  (2004), pp.~137--151.

\bibitem{Berridge2011}
{\sc M.~J. Berridge}, {\em {Calcium signalling and Alzheimer's disease.}},
  Neurochemical research, 36 (2011), pp.~1149--56.

\bibitem{Berridge2014}
{\sc M.~J. Berridge}, {\em {Calcium signalling and psychiatric disease: bipolar
  disorder and schizophrenia.}}, Cell and tissue research, 357 (2014),
  pp.~477--92.

\bibitem{Berridge2000}
{\sc M.~J. Berridge, P.~Lipp, and M.~D. Bootman}, {\em {The versatility and
  universality of calcium signalling.}}, Nature reviews. Molecular cell
  biology, 1 (2000), pp.~11--21.

\bibitem{Bruno2010}
{\sc L.~Bruno, G.~Solovey, A.~C. Ventura, S.~Dargan, and S.~P. Dawson}, {\em
  {Quantifying calcium fluxes underlying calcium puffs in Xenopus laevis
  oocytes}}, Cell Calcium, 47 (2010), pp.~273--286.

\bibitem{Cao2004}
{\sc Y.~Cao, H.~Li, and L.~Petzold}, {\em {Efficient formulation of the
  stochastic simulation algorithm for chemically reacting systems.}}, The
  Journal of chemical physics, 121 (2004), pp.~4059--67.

\bibitem{Cormen2009}
{\sc T.~H. Cormen, C.~E. Leiserson, R.~L. Rivest, and C.~Stein}, {\em
  {Introduction to Algorithms}}, The MIT Press, Cambridge, MA, 3rd~ed., 2009.

\bibitem{DeYoung1992}
{\sc G.~W. {De Young} and J.~Keizer}, {\em {A single-pool inositol
  1,4,5-trisphosphate-receptor-based model for agonist-stimulated oscillations
  in \ca{} concentration.}}, Proceedings of the National Academy of Sciences, 89
  (1992), pp.~9895--9899.

\bibitem{Dickinson2012}
{\sc G.~D. Dickinson, D. Swaminathan and I.~Parker},
{\em {The probability of triggering calcium puffs is linearly related to the number of inositol trisphosphate receptors in a cluster}}, Biophysical Journal, 102 (2012), pp.~1826--1836.

\bibitem{Dickinson2013}
{\sc G.~D. Dickinson and I.~Parker}, {\em {Factors determining the recruitment
  of inositol trisphosphate receptor channels during calcium puffs}},
  Biophysical Journal, 105 (2013), pp.~2474--2484.

\bibitem{Elf2004}
{\sc J.~Elf and M.~Ehrenberg}, {\em {Spontaneous separation of bi-stable
  biochemical systems into spatial domains of opposite phases}}, Systems
  Biology, 1 (2004), pp.~230--236.

\bibitem{Erban:2014:MDB}
{\sc R.~Erban}, {\em From molecular dynamics to {B}rownian dynamics},
  Proceedings of the Royal Society A, 470 (2014), p.~20140036.

\bibitem{Erban2009}
{\sc R.~Erban and S.~J. Chapman}, {\em {Stochastic modelling of
  reaction-diffusion processes: algorithms for bimolecular reactions.}},
  Physical biology, 6 (2009), p.~046001.

\bibitem{Erban:2007:PGS}
{\sc R.~Erban, S.~J. Chapman, and P.~Maini}, {\em A practical guide to
  stochastic simulations of reaction-diffusion processes}.
\newblock 35 pages, available as http://arxiv.org/abs/0704.1908, 2007.

\bibitem{Falcke2003}
{\sc M.~Falcke}, {\em {On the role of stochastic channel behavior in
  intracellular \ca{} dynamics.}}, Biophysical journal, 84 (2003), pp.~42--56.

\bibitem{Flegg2012}
{\sc M.~B. Flegg, S.~J. Chapman, and R.~Erban}, {\em {The two-regime method for
  optimizing stochastic reaction-diffusion simulations.}}, Journal of the Royal
  Society, Interface / the Royal Society, 9 (2012), pp.~859--68.

\bibitem{Flegg2014}
{\sc M.~B. Flegg, S.~J. Chapman, L.~Zheng, and R.~Erban}, {\em {Analysis of the
  two-regime method on square meshes}}, SIAM Journal on Scientific Computing,
  36 (2014), pp.~B561--B588.

\bibitem{Flegg2015}
{\sc M.~B. Flegg, S.~Hellander, and R.~Erban}, {\em {Convergence of methods for
  coupling of microscopic and mesoscopic reaction-diffusion simulations}},
  Journal of Computational Physics,  (2015).

\bibitem{Flegg2013}
{\sc M.~B. Flegg, S.~R\"{u}diger, and R.~Erban}, {\em {Diffusive
  spatio-temporal noise in a first-passage time model for intracellular calcium
  release.}}, The Journal of chemical physics, 138 (2013), p.~154103.

\bibitem{Fraiman2006}
{\sc D.~Fraiman, B.~Pando, S.~Dargan, I.~Parker, and S.~P. Dawson}, {\em
  {Analysis of puff dynamics in oocytes: interdependence of puff amplitude and
  interpuff interval.}}, Biophysical journal, 90 (2006), pp.~3897--3907.

\bibitem{franks2002complexity}
{\sc K.~M. Franks and T.~J. Sejnowski}, {\em {Complexity of calcium signaling
  in synaptic spines}}, BioEssays, 24 (2002), pp.~1130--1144.

\bibitem{Gibson2000}
{\sc M.~A. Gibson and J.~Bruck}, {\em {Efficient exact stochastic simulation of
  chemical systems with many species and many channels}}, The Journal of
  Physical Chemistry A, 104 (2000), pp.~1876--1889.

\bibitem{Gillespie1976}
{\sc D.~T. Gillespie}, {\em {A general method for numerically simulating the
  stochastic time evolution of coupled chemical reactions}}, Journal of
  Computational Physics, 22 (1976), pp.~403--434.

\bibitem{Goldbeter1990}
{\sc A.~Goldbeter, G.~Dupont, and M.~J. Berridge}, {\em {Minimal model for
  signal-induced \ca{} oscillations and for their frequency encoding through
  protein phosphorylation.}}, Proceedings of the National Academy of Sciences
  of the United States of America, 87 (1990), pp.~1461--1465.

\bibitem{Hake2008}
{\sc J.~Hake and G.~T.~Lines},
{\em Stochastic binding of \ca{} ions in the dyadic cleft; continuous versus random walk description of diffusion},
Biophysical Journal, 94 (2008), pp.~4184--4201.

\bibitem{Hao2009}
{\sc Y.~Hao, P.~Kemper, and G.~D. Smith}, {\em {Reduction of calcium release
  site models via fast/slow analysis and iterative
  aggregation/disaggregation}}, Chaos, 19 (2009), pp.~1--13.

\bibitem{Hattne2005}
{\sc J.~Hattne, D.~Fange, and J.~Elf}, {\em {Stochastic reaction-diffusion
  simulation with MesoRD.}}, Bioinformatics (Oxford, England), 21 (2005),
  pp.~2923--4.

\bibitem{holcman2005calcium}
{\sc D.~Holcman, E.~Korkotian, and M.~Segal}, {\em {Calcium dynamics in
  dendritic spines, modeling and experiments.}}, Cell Calcium, 37 (2005),
  pp.~467--475.

\bibitem{Ji2015} {\sc H.~Ji, Y.~Li and S.~H.~Weinberg}, {\em Calcium ion
    fluctuations alter channel gating in a stochastic luminal calcium release
    site model}, IEEE/ACM Transactions on Computational Biology and
  Bioinformatics, PP (2015), 99.

\bibitem{keenersneyd}
{\sc J.~Keener and J.~Sneed}, {\em
  {Mathematical Physiology I: Cellular Physiology}}, Springer, New York, 2nd~ed., 2009.

\bibitem{Koh2006}
{\sc X.~Koh, B.~Srinivasan, H.~S.~Ching and A.~Levchenko},
{\em A 3D monte carlo analysis of the role of dyadic space geometry in spark generation},
Biophysical Journal, 90 (2006), pp.~1999--2014.

\bibitem{Lipkova2011}
{\sc J.~Lipkov\'{a}, K.~C. Zygalakis, S.~J. Chapman, and R.~Erban}, {\em
  {Analysis of Brownian dynamics simulations of reversible bimolecular
  reactions}}, SIAM Journal on Applied Mathematics, 71 (2011), pp.~714--730.

\bibitem{neher2008multiple}
{\sc E.~Neher and T.~Sakaba}, {\em {Multiple roles of calcium ions in the
  regulation of neurotransmitter release}}, Neuron, 59 (2008), pp.~861--872.

\bibitem{Opplestrup:2009:FKM}
{\sc T.~Opplestrup, V.~Bulatov, A.~Donev, M.~Kalos, G.~Gilmer, and B.~Sadigh},
  {\em First-passage kinetic {M}onte {C}arlo method}, Physical Review E, 80
  (2009), p.~066701.

\bibitem{Tyche}
{\sc M.~Robinson}, {\em {Tyche stochastic simulation package}}.
\newblock \url{http://tycheSSA.github.com}.

\bibitem{Robinson}
{\sc M.~Robinson, S.~Andrews, and R.~Erban}, {\em Multiscale reaction-diffusion
  simulations with {S}moldyn}.
\newblock to appear in Bioinformatics, doi: 10.1093/bioinformatics/btv149,
  2015.

\bibitem{rudiger2012termination}
{\sc S.~R{\"u}diger, P.~Jung, and J.-W. Shuai}, {\em Termination of \ca{}
  release for clustered ip3r channels}, PLoS Computational Biology, 8 (2012),
  p.~e1002485.

\bibitem{rudiger2010calcium}
{\sc S.~R{\"u}diger, C.~Nagaiah, G.~Warnecke, and J.~Shuai}, {\em Calcium
  domains around single and clustered IP$_3$ receptors and their modulation by
  buffers}, Biophysical Journal, 99 (2010), pp.~3--12.

\bibitem{rudiger2007hybrid}
{\sc S.~R{\"u}diger, J.~Shuai, W.~Huisinga, C.~Nagaiah, G.~Warnecke, I.~Parker,
  and M.~Falcke}, {\em Hybrid stochastic and deterministic simulations of
  calcium blips}, Biophysical Journal, 93 (2007), pp.~1847--1857.

\bibitem{Rudiger2010}
{\sc S.~R\"{u}diger, J.~W. Shuai, and I.~M. Sokolov}, {\em {Law of mass action,
  detailed balance, and the modeling of calcium puffs}}, Physical Review
  Letters, 105 (2010), 048103.

\bibitem{ruediger2014stochastic}
{\sc S.~R\"{u}diger}, {\em {Stochastic models of intracellular calcium
  signals}}, Phys. Rep., 534 (2014), pp.~39--87.

\bibitem{sherman2001asymptotic}
{\sc A.~Sherman, G.~D. Smith, L.~Dai, and R.~M. Miura}, {\em {Asymptotic
  analysis of buffered calcium diffusion near a point source}}, SIAM J. Appl.
  Math., 61 (2001), pp.~1816--1838.

\bibitem{shuai2009models}
  {\sc J.~W.~Shuai, D.~P.~Yang, J.~E.~Pearson and S.~R\"{u}diger}, {\em An investigation of models of the IP$_3$R channel in Xenopus oocyte},
  Chaos, 19 (2009), 037105.

\bibitem{Smith2009}
{\sc I.~F. Smith and I.~Parker}, {\em {Imaging the quantal substructure of
  single IP3R channel activity during \ca{} puffs in intact mammalian cells.}},
  Proceedings of the National Academy of Sciences of the United States of
  America, 106 (2009), pp.~6404--6409.

\bibitem{Swillens1999}
{\sc S.~Swillens, G.~Dupont, L.~Combettes, and P.~Champeil}, {\em {From calcium
  blips to calcium puffs: theoretical analysis of the requirements for
  interchannel communication.}}, Proceedings of the National Academy of
  Sciences of the United States of America, 96 (1999), pp.~13750--13755.

\bibitem{Tanskanen2007}
{\sc A.~J.~Tanskanen, J.~L.~Greenstein, A.~Chen, S.~X.~Sun and R.~L.~Winslow},
{\em Protein geometry and placement in the cardiac dyad influence macroscopic properties of calcium-induced calcium release},
Biophysical Journal, 92 (2007), pp.~3379--3396.

\bibitem{ullah2012multi}
{\sc G.~Ullah, I.~Parker, D.-O.~D. Mak, and J.~E. Pearson}, {\em Multi-scale
  data-driven modeling and observation of calcium puffs}, Cell Calcium, 52
  (2012), pp.~152--160.

\bibitem{Vais2010}
{\sc H.~Vais, J.~K. Foskett, and D.-O. {Daniel Mak}}, {\em {Unitary Ca(2+)
  current through recombinant type 3 InsP(3) receptor channels under
  physiological ionic conditions.}}, The Journal of General Physiology, 136
  (2010), pp.~687--700.

\bibitem{vanZon:2005:GFR}
{\sc J.~van Zon and P.~ten Wolde}, {\em Green's-function reaction dynamics: a
  particle-based approach for simulating biochemical networks in time and
  space}, Journal of Chemical Physics, 123 (2005), p.~234910.

\bibitem{Wagner1994} {\sc J. Wagner and J. Keizer}, {\em Effects of rapid
    buffers on \ca{} diffusion and \ca{} oscillations}, Biophysical Journal, 67
  (1994), pp.~447--456.

\bibitem{Wang2015} {\sc X. Wang, Y. Hao, S.~H. Weinberg and G.~D. Smith}, {\em
    \ca{}-activation kinetics modulate successive puff/spark amplitude, duration
    and inter-event-interval correlations in a Langevin model of stochastic
    \ca{} release}, Mathematical Biosciences, 264 (2015), pp.~101--107.

\bibitem{weinberg2014influence}
{\sc S.~H. Weinberg and G.~D. Smith}, {\em The influence of \ca{} buffers on
  free [\ca{}] fluctuations and the effective volume of \ca{} microdomains},
  Biophysical Journal, 106 (2014), pp.~2693--2709.

\bibitem{wieder2015exact}
{\sc N.~Wieder, R.~Fink, and F.~von Wegner}, {\em Exact stochastic simulation
  of a calcium microdomain reveals the impact of \ca{} fluctuations on IP$_3$R
  gating}, Biophysical Journal, 108 (2015), pp.~557--567.

\bibitem{Woods2012}
{\sc N.~K. Woods and J.~Padmanabhan}, {\em {Neuronal calcium signaling and
  Alzheimer's disease.}}, Advances in experimental medicine and biology, 740
  (2012), pp.~1193--217.

\bibitem{Yao1995}
{\sc Y.~Yao, J.~Choi, and I.~Parker}, {\em {Quantal puffs of intracellular \ca{}
  evoked by inositol trisphosphate in Xenopus oocytes.}}, The Journal of
  physiology, 482 (1995), pp.~533--553.

\bibitem{zeng2010effect}
{\sc S.~Zeng and W.~R. Holmes}, {\em {The effect of noise on CaMKII activation
  in a dendritic spine during LTP induction}}, J. Neurophysiol., 103 (2010),
  pp.~1798--1808.

\bibitem{zucker2002short}
{\sc R.~S. Zucker and W.~G. Regehr}, {\em {Short-term synaptic plasticity.}},
  Annu. Rev. Physiol., 64 (2002), pp.~355--405.

\end{thebibliography}

\end{document}